\documentclass[pra,twocolumn,nofootinbib,showpacs,eqsecnum]{revtex4}
\usepackage{epsfig,amsmath,amssymb,bm,epsf,graphicx,latexsym,amsfonts}

\pacs{3.67.Lx}

\newcommand{\be}{\begin{equation}}
\newcommand{\ee}{\end{equation}}
\newcommand{\ben}{\begin{eqnarray}}
\newcommand{\een}{\end{eqnarray}}
\newcommand{\bF}{\begin{figure}}
\newcommand{\eF}{\end{figure}}
\newcommand{\dg}{\dagger}

\input{Qcircuit}

\def\ket#1{ | #1 \rangle}
\def\bra#1{{\langle #1 | }}
\def\tr{ {\rm{Tr }}}
\newcommand{\braket}[2]{\mbox{$\langle #1  | #2 \rangle$}}
\newcommand{\proj}[1]{\mbox{$|#1\rangle \!\langle #1 |$}}

\newcommand{\qed}{\vskip-0.4cm\hfill$\square$}

\newtheorem{theorem}{Theorem}
\newtheorem{lemma}{Lemma}

\begin{document}

\title{On the role of entanglement and correlations in mixed-state quantum computation
\footnote{Some of the results in this paper were presented at the
APS March Meeting, 2007, Denver.}}
\author{Animesh Datta}
 \email{animesh@unm.edu}
 \affiliation{Department of Physics and Astronomy, University of New Mexico, Albuquerque, New Mexico 87131-1156, USA}

\author{Guifre Vidal}
 \email{vidal@physics.uq.edu.au }
 \affiliation{School of Physical Sciences, The University of Queensland, QLD 4072, Australia}
\date{\today}

\begin{abstract}
In a quantum computation with pure states, the generation of large
amounts of entanglement is known to be necessary for a speed-up
with respect to classical computations. However, examples of
quantum computations with mixed states are known, such as the DQC1
model [E.~Knill and R.~Laflamme, Phys.\ Rev.\ Lett. {\bf 81}, 5672
(1998)], in which entanglement is at most marginally present, and
yet a computational speed-up is believed to occur. Correlations,
and not entanglement, have been identified as a necessary
ingredient for mixed-state quantum computation speed-ups. Here we
show that correlations, as measured through the operator Schmidt
rank, are indeed present in large amounts in the DQC1 circuit.
This provides evidence for the preclusion of efficient classical
simulation of DQC1 by means of a whole class of classical
simulation algorithms, thereby reinforcing the conjecture that
DQC1 leads to a genuine quantum computational speed-up.
\end{abstract}

\maketitle

\section{Introduction}

    Quantum computation owes its popularity to the realization, more
than a decade ago, that the factorization of large numbers can be
solved exponentially faster by evolving quantum systems than via
any known classical algorithm \cite{Shor}. Since then, progress in
our understanding of what makes quantum evolutions computationally
more powerful than a classical computer has been scarce. A step
forward, however, was achieved by identifying entanglement as a
{\em necessary} resource for quantum computational speed-ups.
Indeed, a speed-up is only possible if in a quantum computation,
entanglement spreads over an adequately large number of qubits
\cite{Jozsa99}. In addition, the amount of entanglement, as
measured by the Schmidt rank of a certain set of bipartitions of
the system, needs to grow sufficiently with the size of the
computation \cite{v03}. Whenever either of these two conditions is
not met, the quantum evolution can be efficiently simulated on a
classical computer. These conditions (which are particular
examples of subsequent, stronger classical simulation results
based on tree tensor networks (TTN) \cite{TTN}) are only
necessary, and thus not sufficient, so that the presence of large
amounts of entanglement spreading over many qubits does not
guarantee a computational speed-up, as exemplified by the
Gottesman-Knill theorem \cite{nc00}.

    The above results refer exclusively to quantum computations with
pure states. The scenario for mixed-state quantum computation is
rather different. The intriguing {\em deterministic quantum
computation with one quantum bit} (DQC1 or `the power of one
qubit') \cite{kl98} involves a highly mixed state that does not
contain much entanglement \cite{dfc05} and yet it performs a task,
the computation with fixed accuracy of the normalized trace of a
unitary matrix, exponentially faster than any known classical
algorithm. This also provides an exponential speedup over the best
known classical algorithm for simulations of some quantum
processes \cite{pklo04}. Thus, in the case of a mixed-state
quantum computation, a large amount of entanglement does not seem
to be necessary to obtain a speed-up with respect to classical
computers.

        A simple, unified explanation for the pure-state and
mixed-state scenarios is possible \cite{v03} by noticing
that the decisive ingredient in both cases is the presence of
{\em correlations}. Indeed, let us consider the Schmidt decomposition of a
vector $\ket{\Psi}$, given by
\begin{equation}
    \ket{\Psi} = \sum_{i=1}^{\chi} \lambda_{i} \ket{i_A}\otimes \ket{i_B},
\end{equation}
where $\braket{i_A}{j_A} = \braket{i_B}{j_B} = \delta_{ij}$
and $\chi$ is the rank of the reduced density
matrices $\rho_A \equiv \tr_B[\proj{\psi}]$ and $\rho_B \equiv
\tr_A[\proj{\psi}$]; and the (operator) Schmidt decomposition of
a density matrix $\rho$ given by \cite{zv04}
\begin{equation}
    \rho = \sum_{i=1}^{\chi^{\sharp}} \lambda^{\sharp}_{i} ~O_{iA} \otimes O_{iB},
\label{eq:Schmidt_rho}
\end{equation}
where $\tr (O_{iA}^{\dagger}O_{jA}) = \tr (O_{iB}^{\dagger}O_{jB})
= \delta_{ij}$. The Schmidt ranks $\chi$ and $\chi^{\sharp}$ are a
measure of correlations between parts $A$ and $B$, with
$\chi^{\sharp} = \chi^2$ if $\rho=\proj{\Psi}$. Let the density
matrix $\rho_t$ denote the evolving state of the quantum computer
during a computation. Notice that $\rho_t$ can represent both pure
and mixed states. Then, as shown in Refs. \cite{v03} and
\cite{TTN}, the quantum computation can be efficiently simulated
on a classical computer using a TTN decomposition if the Schmidt
rank $\chi^{\sharp}$ of $\rho$ according to a certain set of
bipartitions $A:B$ of the qubits scales polynomially with the size
of the computation. In other words, a necessary condition for a
computational speed-up is that correlations, as measured by the
Schmidt rank $\chi^{\sharp}$, grow super-polynomially in the
number of qubits. In the case of pure states (where $\chi =
\sqrt{\chi^{\sharp}}$) these correlations are entirely due to
entanglement, while for mixed states they may be quantum or
classical.

    Our endeavor in this paper is to study the DQC1 model of quantum computation
following the above line of thought. In particular, we elucidate
whether DQC1 can be efficiently simulated with any classical
algorithm, such as those in \cite{v03,TTN} (and, implicitly, in
\cite{Jozsa99}), that exploits limits on the amount of
correlations, in the sense of a small $\chi^{\sharp}$ according to
certain bipartitions of the qubits. We will argue here that the
state $\rho_t$ of a quantum computer implementing the DQC1 model
displays an exponentially large $\chi^{\sharp}$, in spite of it
containing only a small amount of entanglement \cite{dfc05}. We
will conclude, therefore, that none of the simulation techniques
mentioned above can be used to efficiently simulate `the power of
one qubit'.

On the one hand, our result indicates that a large
amount of classical correlations are behind the (suspected)
computational speed-up of DQC1. On the other hand, by showing the
failure of a whole class of classical algorithms to efficiently
simulate this mixed-state quantum computation, we reinforce the
conjecture that DQC1 leads indeed to an exponential speed-up. We note, however, that our result does \emph{not} rule out the possibility that this circuit could be
simulated efficiently using some other classical algorithm.

\section{DQC1 and Tree Tensor Networks (TTN) }
\label{dqcttn}

The DQC1 model, represented in Eq. (\ref{E:circuit}), provides an
estimate of the normalized trace $\tr(U_n)/2^{n}$ of a $n$-qubit
unitary matrix $U_n\in\mathbb{U}(2^n)$ with fixed accuracy efficiently \cite{kl98}.
For discussions on the classical complexity of evaluating the
normalized trace of a unitary matrix, see \cite{dfc05}.

\begin{equation}
\label{E:circuit}
\Qcircuit @C=.5em @R=-.5em {
    & \lstick{\ket{0}\!\bra{0}} & \gate{H} & \ctrl{1} & \meter & \push{\rule{0em}{4em}} \\
    & & \qw & \multigate{4}{U_n} & \qw & \qw \\
    & & \qw & \ghost{U_n} & \qw & \qw \\
    \lstick{\mbox{$I_n/2^n$}} & & \qw & \ghost{U_n} & \qw & \qw \\
    & & \qw & \ghost{U_n} & \qw & \qw \\
    & & \qw & \ghost{U_n} & \qw & \qw \gategroup{2}{2}{6}{2}{.6em}{\{}
}
\end{equation}
This quantum circuit transforms the highly-mixed initial state $\rho_0 \equiv \proj{0}\otimes I_n/2^n$ at time $t=0$ into the final state $\rho_T$ at time $t=T$,
\be
\rho_T = \frac{1}{2^{n+1}}\left(%
\begin{array}{cc}
  I_n & U^{\dg}_n \\
  U_n & I_n \\
\end{array}%
\right),
\label{eq:rhoT}
 \ee
through a series of intermediate states $\rho_t$, $t\in [0,T]$.
The simulation algorithms relevant in the present discussion
\cite{Jozsa99,v03,TTN} require that $\rho_t$ be efficiently
represented with a TTN \cite{TTN} (or a more restrictive
structure, such as a product of $k$-qubit states for fixed $k$
\cite{Jozsa99} or a matrix product state \cite{v03}) at all times
$t\in [0,T]$. Here we will show that the final state $\rho_T$,
henceforth denoted simply by $\rho$, cannot be efficiently
represented with a TTN. This already implies that none of the algorithms in \cite{Jozsa99,v03,TTN} can be used to efficiently simulate the DQC1 model.

Storing and manipulating a TTN requires computational space and
time that grows linearly in the number of qubits $n$ and as a
small power of its rank $q$. The rank $q$ of a TTN is the maximum
Schmidt rank $\chi^{\sharp}_i$ over all bipartitions $A_i:B_i$ of
the qubits according to a given tree graph whose leaves are the
qubits of our system. See \cite{TTN} for details. The key
observation of this paper is that for a {\em typical} unitary
matrix $U_n$, the density matrix $\rho$ in Eq. (\ref{eq:rhoT}) is
such that any TTN decomposition has exponentially large rank $q$.
By {\em typical}, here we mean a unitary matrix $U_n$ efficiently
generated through a (random) quantum circuit. That is, $U_n$ is
the product of poly($n$) one-qubit and two-qubit gates. In the
next section we present numerical results that unambiguously
suggest that, indeed, {\em typical} $U_n$ necessarily lead to TTN
with exponentially large rank $q$.

We notice that the results of the next section do not exclude the
possibility that the quantum computation in the DQC1 model can be
efficiently simulated with a TTN for particular choices of $U_n$.
For instance, if $U_n$ factorizes into single-qubit gates, then
$\rho$ can be seen to be efficiently represented with a TTN of
rank 3, and we can not rule out an efficient simulation of the
power of one qubit for that case. Of course, this is to be
expected, given that the trace of such $U_n$ can be computed
efficiently in the first place.

\section{Exponential growth of Schmidt ranks}

        In this section we study the rank $q$ of any TTN for the final state $\rho$ of the DQC1
circuit, Eq. (\ref{eq:rhoT}). We numerically determine that a
lower bound to such a rank grows exponentially with the number of
qubits $n$.

\begin{figure}[!h]
\begin{center}
 \resizebox{9.2cm}{7cm}{\includegraphics{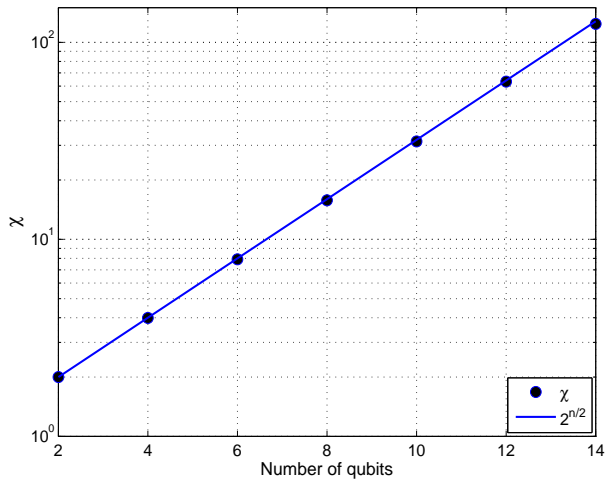}}
  \caption{(Color Online):Lower bound for the operator Schmidt rank $\chi^{\sharp}$ of the DQC1 state for
any equipartition $n/2:n/2$, as given by the Schmidt rank $\chi$
of the pure state in Eq. (\ref{purestate}). The dots are for even
numbers of qubits, and the fit is the line $2^{n/2}$. $\chi$ is
calculated for a pure state obtained by applying $2n$ random
2-qubit gates on the state $\ket{0}^{\otimes n}$. This is evidence
that for a {\em typical} unitary $U_n$, the rank $q$ of any TTN
for the DQC1 state $\rho$ in Eq. (\ref{eq:rhoT}) grows
exponentially with $n$.}
 \label{14qubits}
\end{center}
\end{figure}

The Schmidt rank $\chi$ of a pure state $\ket{\rho_{\phi_A\psi_B}}$
\begin{equation}
\label{purestate}
    \ket{\rho_{\phi_A\psi_B}} \equiv \rho \ket{\phi_A}\ket{\psi_B} =  \sum_{i=1}^{\chi^{\sharp}} \lambda^{\sharp}_{i} ~O_{iA}\ket{\phi_A} \otimes O_{iB}\ket{\psi_B}
\end{equation}
obtained by applying the density matrix $\rho$ onto a product state $\ket{\phi_A}\ket{\psi_B}$ is a lower bound on the
operator Schmidt rank $\chi^{\sharp}$ of $\rho$, i.e.,
$\chi^{\sharp} \geq \chi$. For the purpose of our numerics, we
consider the pure state $U_n\ket{0}^{\otimes n}$.
We build $U_n$ as a sequence of $2n$ random two-qubit
gates, applied to pairs of qubits, also chosen at random. The
random two-qubit unitaries are generated using the mixing algorithm
presented in \cite{ewslc03}. Note that applying $2n$ gates means
that the resulting unitary is efficiently implementable, a
situation for which the DQC1 model is valid. For an even number of
qubits $n$, we calculate the smallest Schmidt rank $\chi$ over all $n/2:n/2$
partitions of the qubits (similar results can be obtained for odd
$n$). The resulting numbers are plotted in Fig (\ref{14qubits}).

        The above numerical results strongly suggest that
the final state $\rho$ in the DQC1 circuit has exponential Schmidt
rank for a {\em typical} unitary $U_n$. We are not able to provide
a formal proof of this fact. This is due to a general difficulty
in describing properties of the set $\mathbb{U}_{qc}(2^{n})$ of
unitary matrices that can be efficiently realized through a
quantum computation. Instead, the discussion is much simpler for
the set $\mathbb{U}(2^n)$ of generic $n$-qubit unitary matrices,
where it is possible to prove that $\rho$ cannot be efficiently
represented with a TTN for a Haar generated $U_n\in
\mathbb{U}(2^n)$, as discussed in the next section. Notice that
Ref. \cite{ell05} claims that random (but efficient) quantum
circuits generate random $n$-qubit gates $U_n\in
\mathbb{U}_{qc}(2^n)$ according to a measure that converges to the
Haar measure in $\mathbb{U}(2^n)$. Combined with the theorem in
the next section, this would constitute a formal proof of the
otherwise numerically evident exponential growth of the rank $q$
of any TTN for the DQC1 final state $\rho$.

%

\section{A formal proof for the Haar-distributed case}

    Our objective in this section is to analyze the Schmidt rank $\chi^{\sharp}$ of
the density matrix $\rho$ in Eq. (\ref{eq:rhoT}) for certain
bipartitions of the $n+1$ qubits, assuming that $U_n \in
\mathbb{U}(2^n)$ is Haar-distributed.

It is not difficult to deduce that for any tree of the $n+1$ qubits, there
exists at least one edge that splits the tree in two parts $A$ and
$B$, with $n_A$ and $n_B$ qubits, where $n_0=\min(n_A,n_B)$
fulfills $n/5 \leq n_0 \leq 2n/5$. In other words, if a rank-$q$
TTN exists for the $\rho$ in Eq. (\ref{eq:rhoT}), then there is a
bipartition of the $n+1$ qubits with $n_0$ qubits on either $A$ or
$B$ and such that the Schmidt rank $\chi^{\sharp} \leq q$. Theorem
\ref{t1}, our main technical result, shows that if $U_n$ is chosen
randomly according to the Haar measure, then the Schmidt rank of
any such bipartition fulfills $\chi^{\sharp} \geq O(2^{n_0})$.
Therefore for a randomly generated $U_n \in \mathbb{U}(2^n)$, a TTN for $\rho$ has
rank $q$ (and computational cost) exponential in $n$, and none of
the techniques of \cite{Jozsa99,v03,TTN} can simulate the
outcome of the DQC1 model efficiently.

%
%

Consider now any bipartition $A:B$ of the $n+1$ qubits, where $A$ and
$B$ contain $n_A$ and $n_B$ qubits, with the minimum $n_0$ of
those restricted by $n/5\leq n_0 \leq 2n/5$. Without loss of
generality we can assume that the top qubit lies in $A$. Actually,
we can also assume that $A$ contains the top $n_A$ qubits. Indeed,
suppose $A$ does not have the $n_A$ top qubits. Then we can use a
permutation $P_n$ on all the $n$ qubits to bring the $n_A$ qubits
of $A$ to the top $n_A$ positions. This will certainly modify
$\rho$, but since
 \be
\left(%
\begin{array}{cc}
  P_n & 0 \\
  0 & P_n \\
\end{array}%
\right)\left(%
\begin{array}{cc}
  I_n & U^{\dg}_n \\
  U_n & I_n \\
\end{array}%
\right)\left(%
\begin{array}{cc}
  P^{T}_n & 0 \\
  0 & P^{T}_n \\
\end{array}%
\right)=\left(%
\begin{array}{cc}
  I_n & V^{\dg}_n \\
  V_n & I_n \\
\end{array}%
\right)
 \ee
where $V_n = P_n U_n P^T_n$ is another Haar-distributed unitary,
we obtain that the new density matrix is of the same form as
$\rho$. Finally, in order to ease the notation, we will assume
that $n_A=n_0$ (identical results can be derived for $n_B=n_0$).
Thus $n/5 \leq n_A \leq 2n/5$.

 We note that
    \be
    \left(%
\begin{array}{cc}
  I_n & U^{\dg}_n \\
  U_n & I_n \\
\end{array}%
\right) = \mathbb{I}_2 \otimes \mathbb{I}_n +  \left(%
\begin{array}{cc}
  0 & 1 \\
  0 & 0 \\
\end{array}%
\right) \otimes U_n^{\dg} + \left(%
\begin{array}{cc}
  0 & 0 \\
  1 & 0 \\
\end{array}%
\right) \otimes U_n,
    \ee
so that if we multiply $\rho$ by the product state
\begin{equation} \label{eq:prod}
 \ket{\phi_{\vec{\alpha}}} \equiv \ket{t,i,j} \equiv \ket{t,i_A}\ket{j_B},
\end{equation}
where $\vec{\alpha} \equiv (t,i,j)$, $t=0,1$; $i=1,\dots d_A$;
$j=1,\dots d_B$, we obtain $\ket{\psi_{\vec{\alpha}}} \equiv \rho
\ket{\phi_{\vec{\alpha}}}$ where
\begin{eqnarray}
\label{pstate}
 \ket{\psi_{\vec{\alpha}}} = \left\{
\begin{array}{c}
\frac{1}{2^{n+1}} (\ket{0,i,j} + \ket{1}\otimes U_n\ket{i,j}) ~~~ \mbox{if} ~t=0\\
\frac{1}{2^{n+1}} (\ket{1,i,j} + \ket{0}\otimes U_n^{\dagger}\ket{i,j}) ~~~ \mbox{if} ~t=1
\end{array} \right.
\end{eqnarray}
This also justifies our choice of the pure state used in the
numerical calculations in the previous section.

Let us consider now the reduced density matrix
    \ben
    \sigma^B_{\vec{\alpha}} &\equiv&
    \tr_A[\proj{\psi_{\vec{\alpha}}}] \nonumber\\
    &=&\frac{1}{2^{n+1}}\left(\ket{j}\bra{j} + \tr_A[U_n\ket{i,j}\bra{i,j} U_n^{\dg}]\right)
    \een
for $t=0$ (for $t=1$, $U_n$ and $U_n^\dagger$ need to be
exchanged). For a unitary matrix $U_n$ randomly chosen according
to the Haar measure on $\mathrm{U}(n)$, $U_n\ket{i,j}$ is a random
pure state on $A\otimes B$. Here, and henceforth $A$ is the space
of the first $n_A$ qubits without the top qubit. It follows from
\cite{hlw06} that the operator
\begin{equation} \label{eq:Q}
    Q = \tr_A [U_n\ket{i,j}\bra{i,j}
U_n^{\dg}]
\end{equation}
has rank $d_A$. Therefore the rank of $\sigma^{B}_{\vec{\alpha}}$
(equivalently, the Schmidt rank $\chi$ of
$\ket{\psi_{\vec{\alpha}}}$) is at least $2^{n_0}$. From Eq.
(\ref{purestate}) we conclude that the Schmidt rank of $\rho$
fulfills $\chi^{\sharp} \geq 2^{n_0} \geq 2^{n/5}$. We can now
collate these results into

\begin{theorem}
 \label{t1}
Let $U_n$ be an $n$-qubit unitary transformation chosen randomly
according to the Haar measure on $\mathrm{U}(2^n)$, and let $A:B$
denote a bipartition of $n+1$ qubits into $n_A$ and $n_B$ qubits,
where $n_0\equiv\min(n_A,n_B)$. Then $n/5\leq n_0\leq 2n/5$ and
the Schmidt decomposition of $\rho$ in Eq. (\ref{eq:rhoT})
according to bipartition $A:B$ fulfills $\chi^{\sharp} \geq
2^{n/5}$.
\end{theorem}

We have seen that we cannot efficiently simulate DQC1 with an
algorithm that relies on having a TTN for $\rho$ with low rank
$q$. However, in order to make this result robust, we need to also
show that $\rho$ canot be well approximated by another
$\tilde{\rho}$ accepting an efficient TTN. We do this in Appendix
\ref{appendixa}.

\section{Conclusions}
\label{disc}

The results in this paper show that the algorithms of
\cite{Jozsa99,v03,TTN} are unable to efficiently simulate a DQC1
circuit. The efficiency of a quantum simulation using these
algorithms relies on the possibility of efficiently decomposing
the state $\rho$ of the quantum computer using a TTN. We have seen
that for the final state of the DQC1 circuit no efficient TTN
exists.

%
%

        It is also interesting to note that the numerics and Theorems 1
and 2 in this paper can be generalized for any fixed polarization
$\tau$, ($0<\tau\leq 1$) of the initial state $\tau\proj{0} +
(1-\tau)\mathbb{I}/2$ of the top qubit of the circuit in Eq
(\ref{E:circuit}), implying that the algorithms of
\cite{Jozsa99,v03,TTN}
 are also unable to efficiently simulate the
power of even the \emph{tiniest} fraction of a qubit.

\section*{Acknowledgements}
AD acknowledges the US Army Research Office for support via
Contract No.~W911NF-4-1-0242 and a Visiting Fellowship from the
University of Queensland, where this work was initiated. GV thanks
support from the Australian Research Council through a Federation
Fellowship.

\appendix

\section{Distribution of the Schmidt coefficients}
\label{appendixa}

In this Appendix we explore the robustness of the statement of
Theorem \ref{t1}. To this end, we consider the Schmidt rank
$\tilde{\chi}^{\sharp}$ for a density matrix $\tilde{\rho}$ that
approximates $\rho$ according to a fidelity $F(O_1,O_2)$ defined
in terms of the natural inner product on the space of linear
operators,
    $$
F(O_1,O_2)\equiv\tr({O_1^{\dg}O_2})\Bigg/{\sqrt{\tr(O_1^{\dg}O_1)}\sqrt{\tr(O_2^{\dg}O_2)}},
    $$
where $F = 1$ if and only if $O_1=O_2$ and $F =
|\braket{\psi_1}{\psi_2}|^2$ for projectors $O_i = P_{\psi_i}$ on
pure states $\ket{\psi_i}$. We will show that if $\tilde{\rho}$ is
close to $\rho$, then $\tilde{\chi}^{\sharp}$ for a bipartition as
in Theorem 1 is also exponential. To prove this, we will require a
few lemmas which we now present.

\begin{lemma}
\label{maxsch}
 Let $\ket{\Psi}$ be a bipartite vector with $\chi$
terms in its Schmidt decomposition,
$$
\ket{\Psi}  = N_{\Psi}\! \sum_{i=1}^{\chi} \lambda_{i}
\ket{i_{A}}\ket{i_{B}},~~ \lambda_{i}\geq \lambda_{i+\!1} \geq 0,
~\sum_{i=1}^{\chi}\lambda_{i}^2 = 1,
$$
where $N_{\Psi} \equiv \sqrt{\braket{\Psi}{\Psi}}$, and let
$\ket{\Phi}$ be a bipartite vector with norm $N_{\Phi}$ and
Schmidt rank $\chi'$, where $\chi' \leq \chi$. Then,
\begin{equation}
\max_{\ket{\Phi}} |\braket{\Psi}{\Phi}| =
N_{\Psi}N_{\Phi}\sqrt{\sum_{i=1}^{\chi'} \lambda_i^{2}}.
\end{equation}
\end{lemma}

{\it Proof: } Let $\mu_i$ denote the Schmidt coefficients of
$\ket{\Phi}$. It follows from Lemma 1 in \cite{vjn00} that $
 \max_{\ket{\Phi}}|\braket{\Psi}{\Phi}| =
 N_{\Psi}N_{\Phi}\sum_{i=1}^{\chi'}\lambda_i\mu_i,
 $
and the maximization over $\mu_i$ is done next. A straightforward
application of the method of Lagrange multipliers provides us with
$\mu_i = c\lambda_i,~i=1,2,\dots,\chi'$ for some constant $c$.
Since $\sum_{i=1}^{\chi'}\mu_i^2=1=c^2\sum_{i=1}^{\chi'}
\lambda_i^2$, $c=1/\sqrt{\sum_{i=1}^{\chi'} \lambda_i^2}.$ Thus,
$$
 \max_{\ket{\Phi}}|\braket{\Psi}{\Phi}| =
c N_{\Psi}N_{\Phi}\sum_{i=1}^{\chi'}\lambda_i^2
 $$
 and the result follows.
\qed

     We will also use two basic results related to majorization theory.  Recall that, by
definition, a decreasingly ordered probability distribution
$\vec{p} = (p_1,p_2,\dots,p_d)$, where $p_{\alpha} \geq
p_{\alpha+1}\geq 0$, $\sum_{\alpha} p_{\alpha}=1$, is {\em
majorized} by another such probability distribution $\vec{q}$,
denoted $\vec{p} \prec \vec{q}$, if $\vec{q}$ is more ordered or
concentrated than $\vec{p}$ (equivalently, $\vec{p}$ is flatter or
more mixed than $\vec{q}$) in the sense that the following
inequalities are fulfilled:
\begin{equation}
    \sum_{\alpha=1}^k p_{\alpha} \leq \sum_{\alpha=1}^k q_{\alpha}~~~\forall~k=1,\dots,d
\end{equation}
with equality for $k=d$.
The following result can be found in Exercise II.1.15 of
\cite{Bhatia}:

\begin{lemma}
    Let $\rho_{\vec{x}}$ and $\rho_{\vec{y}}$ be density matrices with
eigenvalues given by probability distributions $\vec{x}$ and
$\vec{y}$. Let $\sigma(M)$ denote the decreasingly ordered
eigenvalues of hermitian operator $M$. Then
$$
    \sigma(\rho_{\vec{x}} + \rho_{\vec{y}}) \prec \vec{x} + \vec{y}.
$$
\end{lemma}
The next result follows by direct inspection.
\begin{lemma}
Let coefficients $\delta_i$, $1\leq i \leq d$, be such that
$-\delta \leq \delta_i \leq \delta$ for some positive $\delta \leq
1$ and $\sum_i \delta_i = 1$, and consider the probability
distribution $\vec{p}(\{\delta_i\})$,
 $$
    \vec{p}(\{\delta_i\}) \equiv \left(\frac{1}{2} + \frac{1+\delta_1}{2d}, \frac{1+\delta_2}{2d},\cdots, \frac{1+\delta_d}{2d}\right).
 $$
Then
$$
    \vec{p}(\{\delta_i\}) \prec \vec{p}(\{\delta_i^*\}),
$$ where
$$
    \delta_i^* \equiv \left\{
    \begin{array}{c}
    ~~\delta~~~ i \leq d/2 \\
    -\delta ~~~ i > d/2
    \end{array} \right.
$$ and we assume $d$ to be even.
\end{lemma}

Finally, we need a result from \cite{hlw06}:
\begin{lemma}
With probability very close to 1,
\begin{eqnarray}
    \label{hayden}
&&\mathrm{Pr}\Big[(1-\delta)\frac{\Upsilon}{d_A} \leq
Q \leq (1+\delta)\frac{\Upsilon}{d_A} \Big] \nonumber\\
&& \geq1-\left(\frac{10~
d_A}{\delta}\right)^{2d_A}2^{(-d_B\;\delta^2/14 \ln2)} \nonumber\\
&&\geq 1- O\left(\frac{1}{\exp(\delta^2\exp(n))}\right),
\end{eqnarray}
where $d_A = 2^{n_A} = 2^{n_0}$ and $d_B = 2^{n_B} = 2^{n-n_0
+1}$, and the operator $Q$ defined in Eq. (\ref{eq:Q}) is within a
ball of radius $\delta$ of a (unnormalized) projector
$\Upsilon/d_A$ of rank $d_A$ [provided $d_B$ is a large multiple
of $d_A\log d_A/\delta^2$ \cite{hlw06}, which is satisfied for
large $n$, given that $n/5 \leq n_0 \leq 2n/5$].
\end{lemma}

Our second theorem uses the fact that the Schmidt decomposition of
$\rho$ does not only have exponentially many coefficients, but
that these are roughly of the same size.


\begin{theorem}
Let $\rho$, $U_n$, and $A\!:\!B$ be defined as in Theorem
\ref{t1}. If $F(\rho,\tilde{\rho}) \geq 1-\epsilon$, then with
probability $p(\delta,n) = 1 - O(\exp(-\delta^2\exp(n)))$, the
Schmidt rank for $\tilde{\rho}$ according to bipartition $A\!:\!B$
satisfies $\tilde{\chi}^{\sharp} \geq (1-4\epsilon-\delta)
2^{n/5}$.
\end{theorem}

\textit{Proof:}  For any product vector of Eq. (\ref{eq:prod}) we
have
    \ben
    \label{eq:hello}
    |\bra{tij}\rho\tilde{\rho}\ket{tij}|
    &\leq& ~N_{\vec{\alpha}}~\tilde{N}_{\vec{\alpha}}~\sqrt{\sum_{k=1}^{\tilde{\chi}^{\sharp}}(\lambda_k^{ij})^2}
\\\nonumber
    &\leq& N_{\vec{\alpha}}~\tilde{N}_{\vec{\alpha}}\;g(\tilde{\chi}^{\sharp}/d_A),
    \een
where
\begin{equation}
g(x)\equiv \sqrt{\frac{1+(1+\delta)x}{2}}
\end{equation}
and $N_{\vec{\alpha}} \equiv \sqrt{\bra{tij}\rho^2\ket{tij}}$,
$\tilde{N}_{\vec{\alpha}} \equiv
\sqrt{\bra{tij}\tilde{\rho}^2\ket{tij}}$. The first inequality in
(\ref{eq:hello}) follows from Lemma 1, whereas the second one
follows from the fact that the spectrum $\vec{p}$ of
 $$
\rho_B \equiv (N_{\vec{\alpha}})^{-2}\tr_A[\rho\proj{tij}\rho] =
\frac{1}{2}(\proj{j} + Q ),
 $$
where $Q$ has all its $d_A$ non-zero eigenvalues $q_{i}$ in the
interval $2^{-n_0}(1-\delta) \leq q_i \leq 2^{-n_0}(1+\delta)$, is
majorized by $\vec{p}(\{\delta_i^*\})$, as follows from Lemmas 2
and 3. Then,
\begin{eqnarray}
&&1-\epsilon \leq \frac{\tr\rho\tilde{\rho}}{\sqrt{\tr\rho^2}\sqrt{\tr \tilde{\rho}^2}} \label{eq:th1} \nonumber\\
&=& \frac{\sum_{\vec{\alpha}} \bra{\vec{\alpha}}\rho\tilde{\rho}
\ket{\vec{\alpha}}}{ \sqrt{\sum_{\vec{\alpha}'}
\bra{\vec{\alpha}'}\rho^2 \ket{\vec{\alpha}'}
\sum_{\vec{\alpha}''} \bra{\vec{\alpha}''}
\tilde{\rho}^2\ket{\vec{\alpha}''}}
}\label{eq:th2}\nonumber\\
&\leq& g(\tilde{\chi}^{\sharp}/d_A) \frac{\sum_{\vec{\alpha}}
N_{\vec{\alpha}}\tilde{N}_{\vec{\alpha}}}{\sqrt{\sum_{\vec{\alpha}'}
(N_{\vec{\alpha}'})^{2}
 \sum_{\vec{\alpha}''} (\tilde{N}_{\vec{\alpha}''})^{2}} }\nonumber\\
&\leq& g(\tilde{\chi}^{\sharp}/d_A)\nonumber,
\end{eqnarray}
where in the last step we have used the Cauchy-Schwarz inequality,
$|\braket{x}{y}| \leq \sqrt{\braket{x}{x}}\sqrt{\braket{y}{y}}$.
The result of the theorem follows from
$g(\tilde{\chi}^{\sharp}/2^{n_0}) \geq 1-\epsilon $. \qed

\end{document}